\documentclass[12pt,amsfonts]{article}
\usepackage{amsfonts}

\def\t{\textstyle}        
\def\one{1\hskip-.37em 1}

\def\h1{\hskip.5em}
\def\de{\delta}
\def\half{{\textstyle{\frac{1}{2}}}}
\def\wbeta{{\widetilde{\beta}}}
\def\threebytwo{\textstyle{\frac{3}{2}}}
\def\quarter{\textstyle{\frac{1}{4}}}

\def\threequarters{\textstyle{\frac{3}{4}}}

\def\H{{\cal H}}
\def\hk{{\hat\kappa}}

\def\threebytwo{\textstyle{\frac{3}{2}}}

\def\p{\phi}

\def\H{{\cal H}}

\def\g{\gamma}

\def\l{\lambda}
\def\D{{\cal D}}

\def\S{\Sigma'}

\def\E{{\rm I}\hskip-.2em{\rm E}}
\def\ra{\rightarrow}
\def\tint{{\textstyle\int}}
\def\hg{{\hat g}}
\def\hp{{\hat\pi}}
\def\hph{{\hat\phi}}
\def\s{\hskip.08em}

\def\d{\partial}
\def\o{\overline}
\def\a{\alpha}
\def\b{\begin{eqnarray*}}  
\def\e{\end{eqnarray*}}    
\def\bn{\begin{eqnarray}}  
\def\en{\end{eqnarray}}   
\def\<{\langle}
\def\>{\rangle}

\def\no{\nonumber}

\def\k{\kappa}
\def\bx{{\bf x}}

\def\{{\lbrace}
\def\}{\rbrace}
 \title{The Utility of Affine Variables and \\Affine Coherent States}
\author{John R. Klauder\footnote{Email: klauder@phys.ufl.edu}\\
Department of Physics and\\Department of Mathematics\\
University of Florida\\
Gainesville, FL 32611-8440}
\date{ }

\begin{document}
\maketitle
\begin{abstract}
Affine coherent states are generated by affine kinematical variables much like
canonical coherent states are generated by canonical kinematical variables.  Although
all classical and quantum formalisms normally entail canonical variables, it is shown
that affine variables can serve equally well for many classical and quantum studies. This
general purpose analysis provides tools to discuss two major applications: (1) the completely successful quantization of a nonrenormalizable scalar quantum field theory by affine techniques, in complete contrast to canonical techniques which only
offer triviality; and (2) a formulation of the kinematical portion of quantum gravity that
favors affine kinematical variables over canonical kinematical variables, and which generates a framework in which a
favorable analysis of the constrained dynamical issues can take place. All this is possible
because of the close connection between the affine and the canonical stories, while the few distinctions can be used to advantage when appropriate.
\end{abstract}
\section*{Background and Introduction}
The standard kinematical variables for quantum mechanics are the self-adjoint Heisenberg pair
$P$ and $Q$ that satisfy the canonical commutation relation $[Q,P]=i\s\hbar\s\one$, each with a spectrum
that covers the entire real line.
These operators are used throughout quantum mechanics, and, for purposes of the present article, one of their important roles lies in their
use in the formation of canonical coherent states, which, for all $(p,q)\in{\mathbb R}^2$, are given
as abstract vectors $|p,q\>$ in an abstract Hilbert space $\frak H$ by
  \b |p,q\>\equiv e^{\t i(p\s\s Q-i\s qP)/\hbar}\,|\eta\> \;.\e
  Here the unit vector $|\eta\>$ is called the fiducial vector, and typically $|\eta\>=|0\>$ where
  $(\omega\s Q+iP)\s|0\>=0$, with $\omega=1$ a common choice and assuming for convenience that $Q$ and $P$ have the same dimension. In a Schr\"odinger representation, where $Q\s|x\>=x\s|x\>$, it follows that
     \b  \<x|0\>=\psi_0(x)=(\pi\hbar)^{-1/4}\,e^{\t- x^2/2\hbar}\;.\e
  The operators $P$ and $Q$ are required to be self adjoint, and thus the coherent states $|p,q\>$
  all have unit norm. Generally, the dependence of $|p,q\>$ on $|\eta\>$ is left implicit.
  The canonical coherent states are strongly continuous in their labels $(p,q)$, and
  in addition, the coherent states admit a resolution of the unit operator in the Hilbert space $\frak H$ given by \cite{thesis,the}
    \b \one=\int |p,q\>\<p,q|\,dp\s\s dq/2\s\pi\s\hbar\;, \e
   which holds for arbitrary $|\eta\>$ when integrated over the whole phase space ${\mathbb R}^2$. More precisely, this resolution of unity holds in the weak sense
      \b \<\phi|\psi\>=\int\<\phi |p,q\>\<p,q|\psi\>\,dp\s\s dq/2\s\pi\s\hbar \e
      for arbitrary $|\phi\>$ and $|\psi\>$ in $\frak H$, as well as in the strong sense.

      The affine coherent states are also based on two kinematical operators $Q$ and $D$ which satisfy the
      affine commutation relation $[Q,\s D]=i\s\hbar\s Q$ \cite{aff22}.  It follows \cite{russia,rr} that there are {\it three}
      inequivalent representations of the affine commutation relations: one with $Q>0$, one with $Q<0$, and one with $Q=0$. We will be primarily interested in the first two representations, and for the present, we assume that $Q>0$ is our choice. For convenience, let us assume that $Q$ is dimensionless,
      while, of necessity, $D$ has the dimensions of $\hbar$. In that case, the affine coherent states are given by
        \b |p,q\>\equiv e^{\t i\s p\s\s Q/\hbar}\,e^{\t-i\s\ln(q)\s D/\hbar}\,|\eta\>\;, \e
        for all $(p,q)\in(\mathbb R,{\mathbb R}^+)$, i.e., $q>0$. In the present case, it is common to choose $|\eta\>$ as the solution to the equation $[\s Q-1+iD/(\beta\hbar)]\s|\eta\>=0$, from which it follows that $\<\eta|\s Q\s|\eta\>=1$ and $\<\eta|\s D\s|\eta\>=0$. In certain situations it is useful to regard ${\widetilde\beta}\equiv \beta\s\hbar$ and $\hbar$ as two independent variables rather than $\beta$ and $\hbar$. In this view, it follows that
         \b \<\eta|\s Q^2\s|\eta\>=1+1/(2\beta)=1+\hbar/(2{\widetilde\beta})=1+O(\hbar)\;,\e
         and indeed, $\<\eta|\s Q^p\s|\eta\>=1+O(\hbar)$, $p\ge1$.
          The Schr\"odinger representation for $|\eta\>$ is given (for $x>0$) by
          \b \<x|\eta\>=\psi_\eta(x)= N\,x^{\beta-\half}\,e^{\t-\beta\s x}\;. \e
       The affine coherent states are continuous in $(p,q)$, and
        also admit a resolution of unity in the form
          \b \one=\int |p,q\>\<p,q|\,dp\s\s dq/[2\s\pi\s\hbar\s C]\;, \e
          where $C \equiv \<\eta|\s Q^{-1}\s|\eta\>=[1-1/(2\beta)]^{-1}<\infty$, which limits $\beta$ so that $\beta>1/2$. Again, the resolution of unity holds in the weak and strong sense.

          Observe that the Heisenberg operators $P$ and $Q$ do not form a Lie algebra by themselves, while  the affine operators $Q$ and $D$ do form a Lie algebra. Indeed, modulo linear transformations, the affine variables form the elements of the only two-parameter, non-Abelian Lie algebra. The affine group gets its name from its realization as the
          set of affine transformations of the real line given by $x\ra x'=a\s x+b$, where $a\not=0$ and $b\in{\mathbb R}$; indeed, the affine group is often called the ``$a\s x+b$'' group!

          \subsection*{Connection between the canonical and affine algebras}
          On the surface, it seems like the canonical
          operator pair and the affine operator pair would have little in common, but that conclusion would be incorrect. Consider the Heisnberg commutator $[Q,P]=i\s \hbar\one$ and multiply both sides by $Q$ leading to $[Q,P]\s Q=i\s\hbar\s Q$. Next bring the $Q$ inside the bracket giving  $[Q,P\s Q]=i\s\hbar\s Q$, and then symmetrize the factor $P\s Q$ (allowed because the difference commutes with $Q$), which leads to
          $[Q,(Q\s P+P\s Q)/2]=i\s\hbar\s Q$, or, finally,  we arrive at
             \b [Q, D]=i\s\hbar\s Q\;,\hskip3em D\equiv \half(Q\s P+P\s Q)\;.  \e
          In short, the affine commutation relation is---or surely seems to be---a consequence of the
          canonical commutation relation. Since the spectrum of each of the Heisenberg variables $Q$ and $P$ is the whole real line, the so-derived affine commutation relation is reducible because it contains
          $Q>0$, $Q<0$, and also $Q=0$ in a sense. The range of $P$ is the real line, so the range of $D$ is also the real line; in fact, that range is doubly covered: once when $Q>0$, and once when $Q<0$.
          We can also recover the canonical commutation relation
          from the affine commutation relation as follows: Act on $Q$ and $D$ by the unitary transformation
          $U[\gamma]\equiv \exp(-i\s\gamma P/\hbar)$, which leads to
             \b U[\gamma]^\dag\s Q\s\s U[\gamma]=Q+\gamma\s\one\;,\hskip2em U[\gamma]^\dag\s D\s\s U[\gamma]=D+\gamma\s P\;,\e
             The same unitary transformation acting on the affine commutation relation leads to
             \b [Q+\gamma\s\one, D+\gamma\s P]=i\s\hbar\s (Q+\gamma\one) \;,\e
             or $[Q,D+\gamma\s P]=i\s\hbar\s(Q+\gamma\s\one)$. Division by $\gamma$ followed by the limit $\gamma\ra\infty$ leads to $[Q,P]=i\s\hbar\s\one$ as desired! Interestingly enough,
             one can derive the canonical algebra with a full real-line spectrum for both $Q$ and $P$ by initially
             starting with the affine algebra with $Q>0$. This property holds because after the given unitary
             transformation acting on $Q$ leads to $Q+\gamma\s\one$, it follows that the spectrum of $Q$
             now satisfies $Q>-\gamma\s\one$. When $\gamma\ra\infty$, the spectrum of $Q$ becomes the whole real line.
              We clearly see that the
             canonical algebra and the affine algebra are quite closely related; for a personal historical
             remark on that relation see \cite{kla1953}.

             Another version of how one kinematical system
      passes to the other kinematical system arises for the coherent state overlap function for the two cases under study. For the canonical coherent states defined with the fiducial vector chosen as the solution of $(Q+iP)\s|0\>=0$, it follows that
               \b  \<p',q'|p,q\>=e^{\t i(q'p-p'q)/2\hbar-[(p'-p)^2+(q'-q)^2]/4\hbar }\;. \e
               For the affine coherent states, with the fiducial vector chosen to satisfy the standard equation above, the affine coherent state overlap function reads
                \b \<p',q'|p,q\>=[\s\s\half\s\sqrt{q'/q}+\half\s\sqrt{q/q'}+i\half\s\sqrt{q'\s q}\s(p'-p)/(\beta\hbar)\s\s]^{-2\beta}\;.\e
                In this latter relation, $q'$ and $q$ are both positive. However, if we let $q'\ra q'+\gamma$ and
                $q\ra q+\gamma$, and also choose $\beta=\gamma^2$, then, as $\gamma\ra\infty$, it follows that
                the affine coherent state overlap function converges to the canonical coherent state overlap up to a rescaling of the variables to restore proper dimensions and
                an overall phase that can readily be absorbed into a redefinition of the canonical coherent states if desired.

\section*{Comments on Classical and Quantum \\Mechanics}
The Hamiltonian formulation of classical mechanics is remarkable in its own right, and it is natural as a formalism
to bridge the gap between classical and quantum mechanics. The classical action functional serves as an ideal
starting point for our discussion, and for a single degree of freedom the action is given by
  \b A=\tint[\s p\s\s{\dot q}-H(p,q)\s]\,dt\;. \e
  Stationary variation of $A$ leads to Hamilton's equations of motion given by
    \b {\dot q}=\d H(p,q)/\d p\equiv \{q,H\}\;,\hskip3em {\dot p}=-\d H(p,q)/\d q\equiv \{p,H\}\;, \e
    where we have introduced the Poisson bracket
      \b \{A,B\}\equiv \frac{\d A}{\d q}\s\frac{\d B}{\d p}-\frac{\d A}{\d p}\s\frac{\d B}{\d q}\;. \e
      As is well known, the Hamiltonian formalism retains form invariance under the wide class of canonical transformations from one set of phase space coordinates to another. In particular, to pass between
      the phase space coordinates $(p,q)$ and $({\o p},{\o q})\equiv({\o p}(p,q),{\o q}(p,q))$, the connecting relation is usually given by
        \b {\o p}\s\s d{\o q}=p\s\s dq+dF({\o q},q)\;, \e
        where $F$ is known as the generator of the transformation. One such transformation that will be of interest to us, is
           \b {\o p}\s\s d{\o q}=(p\s\s q)\s\s d\ln(q)=(p\s\s q)\s\s q^{-1}\s dq=p\s\s dq \;,\e
        for which $F=0$. The new variables are ${\o p}\equiv p\s\s q\equiv d$ and ${\o q}\equiv \ln(q)\equiv s$. Of course, for these variables to be well defined, it is necessary that $q>0$. However, this boundary can easily
        be shifted simply by using $d\equiv p\s (q+\gamma)$ and $s\equiv \ln(q+\gamma)$, where now $q>-\gamma$. If $\gamma$ is $10^{137}$ in suitable units, then for all practical purposes one would be secure in treating $q$ as unbounded. As constructed, $(d,s)$ are canonical phase-space coordinates for which
        $\{s,d\}=1$. It is also interesting to note that $\{q,d\}=q$, which is clearly the classical version of the affine commutation relation. The variables $(d,q)$ are not canonical variables, but the affine Poisson bracket does arise
        from the canonical Poisson bracket $\{q,p\}=1$ simply by multiplying both sides by $q$ leading to
          \b q=\{q,p\}\s q =\{q,p\s\s q\}=\{q,d\}\;, \e
          as desired.

          Based on the fundamental Poisson bracket $\{q,d\}=q$, it is even possible to reformulate classical mechanics completely in
          terms of affine variables. Besides the usual Hamiltonian $H(p,q)$, let us introduce $H'(d,q)\equiv H(p,q)$
          and treat $d$ and $q$ as new, independent variables.
             To derive the equations of motion, we note that
              \b {\dot q}\hskip-1.3em&&=\{q,H\}=\frac{\d H}{\d p}=q\s\s\frac{\d H'}{\d d}\;,\no\\
             {\dot d}\hskip-1.3em&&=\{d,H\}=p\s\frac{\d H}{\d q}-q\s\frac{\d H}{\d p}=\frac{d}{q}\s\frac{\d H'}{\d q}-q^2\s\frac{\d H'}{\d d}\;.\e
             Of course, these equations do not have the familiar symmetry of the usual Hamiltonian equations, but they
             are nevertheless correct.

\subsection*{Toy model: classical study}
             Let us discuss one simple example of a dynamical system where $q>0$. Consider the action functional given by
              \b A=\tint[\s -q\s{\dot p}-q\s p^2\s]\,dt\;,\e
              which we regard as a toy model of classical gravity for which $q(t)>0$ represents the metric $g_{\mu\s\nu}(x)$ with its signature constraints and $p(t)$ represents (minus) the Christoffel symbol $\Gamma^\a_{\beta\s\gamma}(x)$; see, e.g., \cite{ak}. The equations of motion that follow from this toy model
              are given by
                \b {\dot q}=2\s p\s q\;, \hskip3em {\dot p}=-p^2\;, \e
                which have the general solutions
                 \b p(t)=p_0(1+p_0\s t)^{-1}\;, \hskip3em q(t)=q_0\,(1+p_0\s t)^2\;, \e
                 where $(p_0,q_0)\in({\mathbb R},{\mathbb R}^+)$ are initial values at $t=0$.
                 Clearly, {\it every} solution with positive energy, i.e., $E_0=q_0\s p_0^2>0$, exhibits
                 a singularity at $t=-1/p_0$; only if $p(t)=p_0=0$ and $q(t)=q_0>0$ are there no singularities.

                 Let us see if quantum corrections to the clasical equations can possibly eliminate these singularities.

             \section*{The Quantum/Classical Connection}
             When directly quantizing a classical canonical system---Note: a direct canonical quantization, as used in this article, involves promoting $p\ra P$, $q\ra Q$, and $H(p,q)\ra \H=H(P,Q)$ modulo natural factor ordering---the usual
             rules state that Cartesian coordinates must be used. This rule may be understood on the basis of
             the Weak Correspondence Principle \cite{wcp}, which states for canonical systems that the quantum and classical Hamiltonians are related through the expression
               \b H(p,q)=\<p,q|\s\H(P,Q)\s|p,q\>=\<0|\s\H(P+p,Q+q)\s|0\>=\H(p,q)+O(\hbar;p,q)\;.\e
               For example, if $H(p,q)=p^2+q^4$, then
               \b \<0|\s[\s(P+p)^2+(Q+q)^4\s]\s|0\>=p^2+q^4+O(\hbar;p,q)\;,\e
                where in this case $O(\hbar;p,q)=6\s q^2\s\<0|Q^2|0\>+\<0|[P^2+Q^4]|0\>=(\half+3\s q^2)\hbar +\threequarters
                \hbar^2$. These expressions link the quantum Hamiltonian to the functional form of the classical Hamiltonian, but it is not clear what makes these coordinates ``Cartesian''. To
               address that issue, we recall, for the given choice of canonical coherent states, the facts that
                \b \theta(p,q)\hskip-1.3em&&\equiv i\s\hbar\<p,q|\s\s d|p,q\>=\half\s(p\s\s dq-q\s\s dp)\;,\no\\
                 d\sigma^2(p,q)\hskip-1.3em&&\equiv2\s\hbar[\s\s \|\s\s d|p,q\>\|^2-|\<p,q|\s\s d|p,q\>|^2\s]=dp^2+dq^2\;, \e
               which are classical expressions, derived from the quantum theory, that offer a familiar one form and a (Fubini-Study)
                Cartesian metric to the
                classical theory. In alternative canonical coordinates, for which the coherent states transform as $|{\o p},{\o q}\>\equiv |p({\o p},{\o q}),q({\o p},{\o q})\>=|p,q\>$, the metric above would generally not be Cartesian and
                thus a direct quantization procedure would generally lead to an incorrect result. For example, it is possible to choose canonical coordinates $({\o p},{\o q})$ such that the Hamiltonian
                $p^2+q^4={\o p}^2$. When that is the case, $P^2+Q^4\not={\o P}^2$ for the simple reason that one operator has a discrete spectrum while the other has a continuous spectrum.

                There is an almost parallel story that applies to the affine variables and the chosen form of the affine coherent states. In this case, we claim that promoting $d\ra D$ and $q\ra Q$, even when $q>0$ as well as $Q>0$, also leads to a
                satisfactory direct quantization rule; this rule holds because
                 \b H'(p\s q,q)=\<p,q|\s\H'(D,Q)\s|p,q\>\hskip-1.3em&&=\<\eta|\s \H'(D+p\s q\s Q,q\s Q)\s|\eta\>\no\\
                 &&=\H'(p\s q,q)+O(\hbar;p,q)\;.\e
                 This relation uses the property that $\<\eta|\s Q^p\s|\eta\>=1+O(\hbar)$, which,
                 as argued previously, follows from the equation that defines
                 the affine fiducial vector as a function of $\wbeta$ and $\hbar$. For example, if $H'(d,q)=d^2+q^4$, then
                   \b \<p,q|\s[\s D^2+Q^4\s]\s|p,q\>=\<\eta|\s[ (D+p\s q\s Q)^2+(q\s Q)^4\s]\s|\eta\>=d^2+q^4+O(\hbar;p,q)\;, \e
                   where, now $O(\hbar;p,q)=d^2\<\eta|[Q^2-1]|\eta\>+q^4\<\eta|[Q^4-1]|\eta\>+\<\eta|D^2|\eta\>$.
                The affine coherent states also define a one form and a metric given by
                \b \theta(p,q)\hskip-1.3em&&\equiv i\s\hbar\<p,q|\s\s d|p,q\>=-q\s\s dp\;,\no\\
                 d\sigma^2(p,q)\hskip-1.3em&&\equiv 2\s\hbar[\s \|\s\s d|p,q\>\|^2-|\<p,q|\s\s d|p,q\>|^2\s]=\wbeta^{-1}\s q^2\s dp^2+\wbeta\s q^{-2}\s dq^2\;, \e
                 The one form is again familiar, while now the (Fubini-Study) metric describes a two-dimensional space with
                 a constant negative scalar curvature:  $-2/\wbeta$. This space seems quite distinct from the flat
                 space given by the canonical coherent states, but that can be remedied by replacing $q$
               by $q+\gamma$ leading to the metric $\wbeta^{-1}\s (q+\gamma)^2\s dp^2+\wbeta\s (q+\gamma)^{-2}\s dq^2$, which tends to a flat space with a Cartesian  metric as $\beta=\gamma^2\ra\infty$.

               It is noteworthy that a direct affine quantization of a classical Hamiltonian, i.e., $H'(d,q)\ra
               \H'=H'(D,Q)$ modulo natural factor ordering,  expressed in terms of affine variables $d$ and $q$, where $d$ is constructed as $d=p\s q$ from {\it Cartesian coordinates $p$ and $q$}, leads to a valid quantization just as does a direct quantization of Cartesian canonical coordinates. As with the canonical story, a general change of affine coordinates would normally yield a Hamiltonian function that upon direct quantization would lead to incorrect results.

\subsection*{Toy model: quantum study}
Let us reexamine the toy model of gravity introduced above. We start with a direct canonical quantization in which the Hamiltonian becomes $H=p\s\s q p\ra\H=P\s Q\s P$, and also introduce canonical coherent states $|p,q\>$ as defined above. In that case, the extended  classical Hamiltonian [i.e., extended by possible terms $O(\hbar)$, and often referred to as a ``symbol''] is given by
                  \b H(p,q)=\<p,q|\s P\s Q\s P\s|p,q\>\hskip-1.3em&&=\<0|\s (P+p)(Q+q)(P+p)\s|0\>\no\\
                  &&=q p^2+q\s\<0|P^2|0\>\no\\
                  &&\equiv q p^2+a^2 q\;,  \e
                  where $a^2=\half\hbar$.   Thus, the extended equations of motion are given by
                   \b {\dot q}=2\s p\s q\;, \hskip3em {\dot p}=-(p^2+a^2)\;, \e
                   which lead to  the solutions
                  \b p(t)=a\cot(a\s (t+\a))\;,\hskip3em q(t)=(E_0/a^2)\s \sin(a\s (t+\a))^2 \;, \e
                  where  $E_0=q_0(p_0^2+a^2)>0$, and these solutions exhibit
                   singularities for {\it all} energies (since $q_0>0$ is required). Thus a direct canonical
                   quantization of this toy model has not helped in removing singularities.

                Next, let us reexamine the same toy model using a direct affine quantization in which the Hamiltonian becomes $H=p\s q p=d\s q^{-1} d=H'\ra \H'=D\s Q^{-1} D$, and also introduce  affine coherent states instead of canonical coherent states. For affine coherent states, recall that $\<\eta|D|\eta\>$=0 and $\<\eta|Q|\eta\>=1$, and moreover, $\<\eta| Q^p|\eta\>=1+O(\hbar)$, for all $p\ge-1$; this property  implies that $Q$ (and $q$) are dimensionless, while the dimensions of $P$ (and $p$) are those of $\hbar$. In this case, the extended classical Hamiltonian is given by
                    \b H(p,q)=\<p,q|\s DQ^{-1}D\s|p,q\>\hskip-1.3em&&=\<\eta|\s (D+p\s q\s Q)(q\s\s Q)^{-1}(D+p\s q\s Q)\s|\eta\>\no\\
                      &&=q p^2+ \<\eta|\s DQD\s|\eta\>\,q^{-1}\no\\
                      &&\equiv q p^2+\hbar^2\s C\s q^{-1}\;, \e
                     for some dimensionless constant $C>0$. It is already clear from the form of the extended Hamiltonian that {\it all} solutions to the extended equations of motion with finite energy are nonsingular because
                      \b E_0=q(t)\s p(t)^2+\hbar^2\s C\s q(t)^{-1}\ge \hbar^2\s C\s q(t)^{-1}\;. \e
                      In this case the extended equations of motion are given by
                        \b  {\dot q}=2\s p\s q\;, \hskip3em {\dot p}=-p^2+\hbar^2 C\s q^{-2}\;, \e
                        and the new solutions are given by
                    \b q(t)=4E_0[ (t+\a)^2+\hbar^2\s C/4E_0^2]\;,\hskip3em p(t)=\frac{(t+\a)}{{(t+\a)^2+\hbar^2\s C/4E_0^2}}\;,\e
                    where $E_0=q_0p_0^2+\hbar^2\s C/q_0>0$.

                  It is noteworthy that using a direct affine quantization and affine coherent states, instead of a direct canonical quantization and canonical coherent states, leads to an extended classical Hamiltonian that eliminates singularities in solutions to the extended classical equations of motion, singularities that were present in the usual classical theory.

               \section*{Coherent State Path Integration}
               Here we offer a brief account of path integral quantization using both canonical coherent states and affine coherent states; for more details, see \cite{book}.  For both cases, the propagator is given by
                  \b &&\hskip-2em\<p'',q''|\s e^{\t-i\H\s T}\s|p',q'\>\no\\
                  &&\hskip0em={\cal M}\int e^{\t (i/\hbar)\tint[\xi\s p\s{\dot q}-\zeta\s q{\dot p}-H(p,q)\s]\,dt}\,{\cal D}p\s {\cal D}q\no\\
                  &&=\lim_{N\ra\infty}\int \Pi_{n=0}^N\<p_{n+1},q_{n+1}|p_n,q_n\>\,
                   \;e^{\t-(i\epsilon/\hbar)\Sigma_{n=0}^N H(p_{n+1},q_{n+1};p_n,q_n)}\no\\ &&\hskip11em\times\Pi_{n=1}^N\s dp_n\s dq_n/2\pi\hbar[C]\;. \e
                  Here, $T=\epsilon\s(N+1)$, $(p'',q'')=(p_{N+1},q_{N+1})$, $(p',q')=(p_0,q_0)$, $[C]=1$,
                  $\xi=\zeta=1/2$ for the
                  canonical case, $[C]=\<\eta|Q^{-1}|\eta\>$, $\xi=0$, $\zeta=1$ for the affine case, and
                  for both cases $H(p,q;r,s)=\<p,q|\H\s|r,s\>/\<p,q|r,s\>$ for the relevant coherent states.

                  Other versions of coherent state path integral quantization also exist. One important
                  formulation involves a Wiener measure regularization and takes the form
                   \b  &&\hskip-1.3em\<p'',q''|\s e^{\t-i\H\s T}\s|p',q'\>={\cal M}\int e^{\t (i/\hbar)\tint[\xi\s p\s{\dot q}-\zeta\s q{\dot p}-h(p,q)\s]\,dt}\,{\cal D}p\s {\cal D}q\no\\
                   &&=\lim_{\nu\ra\infty}\,\int e^{\t (i/\hbar)\tint[\xi p\s\s {\dot q}-\zeta q\s {\dot p} -h(p,q)\s]\,dt}\;e^{\t -(1/2\nu)\tint [d\sigma^2(p,q)/(dt)^2]\s dt}\no\\
                   &&\hskip6em\times\Pi_t\s dp(t)\s dq(t)/2\pi\hbar[C]\no\\
                   &&=\lim_{\nu\ra\infty}\,2\pi\hbar[C]\,e^{\nu\s T/2\hbar}\int e^{\t (i/\hbar)\tint[\xi p\s\s dq-\zeta q\s dp -h(p,q)\s dt]}\;d\mu_W^\nu(p,q)\;;\e
                   in the present case, $h(p,q)$ is implicitly defined by the relation
                      \b \H=\int h(p,q)\,|p,q\>\<p,q|\,dp\s\s dq/2\pi\hbar[C]\;, \e
                      for the relevant coherent states, $d\sigma^2(p,q)$ is the relevant metric, and $\mu^\nu_W(p,q)$ is a  two-dimnensional Wiener measure,  with diffusion constant $\nu$,  which is pinned so that  $(p(T),q(T))$ $=(p'',q'')$ and
                      $(p(0),q(0))=(p',q')$. Even for these Brownian motion paths, terms such as $\tint p\s\s dq$ are well-defined stochastic integrals, taken here in the Stratonovich formulation  (i.e., the mid-point rule). Note well, that unlike the lattice regularization of the former
                      coherent state path integral, the latter formulation that involves a Wiener measure regularization involves {\it continuous} phase-space paths for all $\nu<\infty$, and enjoys a well-defined functional integral over these continuous paths for all $\nu<\infty$.

                      \section*{Ultralocal Field Quantization}
                      We next take a big leap from systems with  a single degree of freedom to fields, which are systems
                      with an infinite number of degrees of freedom. Although field problems are technically more complicated than the the simpler ones studied so far, we will
                      endeavor to frame our discussion in terms that make close contact with what has
                      been already presented.

                      \subsection*{Canonical quantization of ultralocal models}
                      Consider the classical phase-space action functional
                      \b A_0=\tint\{\s[\pi(t,\bx)\s{\dot \p}(t,\bx) -\half\s[\s\pi(t,\bx)^2+m_0^2\s\p(t,\bx)^2\s]\s\}\,dt\s d\bx\;,\e
                      which describes the free ultralocal scalar field model. This model differs from a relativistic  free theory by the absence of the term $[\nabla\p(t,\bx)]^2$, and therefore the
                      temporal behavior of the field at one point $\bx$ is independent of the temporal behavior of the field at any
                      point $\bx'\not=\bx$.  Direct canonical quantization of the classical Hamiltonian,
                      i.e., $\pi(\bx)\ra\hp(\bx)$ and $\p(\bx)\ra\hph(\bx)$ where  $H(\pi,\p)\ra\H=H(\hp,\hph)$, leads to an infinite ground state energy, so it is traditional to proceed differently. Formally speaking, the classical Hamiltonian density
                      is first reexpressed as
                       \b \half[\pi(\bx)^2+m_0^2\s\p(\bx)^2]=\half[m_0\p(\bx)-i\s\pi(\bx)][m_0\p(\bx)+i\s\pi(\bx)]\;,\e
                       and only then it is quantized directly. This procedure leads to the quantum Hamiltonian
                       \b \H\hskip-1.3em&&=\half\int\{[m_0\hph(\bx)-i\s\hp(\bx)][m_0\hph(\bx)+i\s\hp\bx)]\,d\bx\no\\
                       &&=\half\int[\hp(\bx)^2+m_0^2\s\hph(\bx)^2-\hbar\s m_0\s\delta(0)]\,d\bx\no\\
                       &&\equiv\half\int\s:[\hp(\bx)^2+m_0^2\s\hph(\bx)^2]:\,d\bx\;, \e
                       a result which leads to conventional normal ordering symbolized by $:\,:\s$.

                       If we regularize this expression by a finite, $s$-dimensional, hypercubic, spatial lattice,  then the Hamiltonian operator
                       reads
                       \b \H_0=\half\S_k[-\hbar^2 \s a^{-2s}\d^2/\d\p_k^2+m^2_0\s\p_k^2-\hbar\s m_0\s a^{-s}]\,a^s\;. \e
                       Here $a>0$ is the lattice spacing, $a^s$ is an elementary cell spatial volume, and we
                      have used the fact that on the lattice $\hp(\bx)\ra-i\hbar\s a^{-s}\d/\d\p_k$ and $\hph(\bx)\ra \p_k$, where $k=\{k_1,\ldots,k_s\}$, $k_j\in{\mathbb Z}$, labels
                       points on the spatial lattice.  The ground state of this Hamiltonian is just the
                       product of a familiar Gaussian ground state,
                         \b \psi_k(\p_k)=(m_0\s a^s/\pi\hbar)^{1/4}\, e^{\t - m_0\p_k^2\s a^s/2\hbar}\e
                       for a large number of independent, one-dimensional harmonic oscillators, and thus
                      the characteristic functional (i.e., the Fourier transform) of the
                      ground-state distribution is given by
                        \b C_0(f)\hskip-1.3em&&=\lim_{a\ra0}N_0\int e^{\t (i/\hbar)\S_k\s f_k\s\p_k\, a^s-(m_0/\hbar)\S_k\p_k^2\, a^s}\,\Pi'_k d\p_k\no\\
                            &&= {\cal N}_0\int e^{\t (i/\hbar)\tint f(\bx)\s\p(\bx)\,d\bx-(m_0/\hbar)\tint \p(\bx)^2\,d\bx}\,\Pi'_\bx d\p(\bx)\no\\
                            &&=e^{\t -(1/4\hbar m_0)\tint f(\bx)^2\,d\bx}\;. \e
                     Here, in the last two lines the continuum limit has been taken (with a formal
                     version in the second line)
                     in which $a\ra0$, $L$, the number of sites on each edge diverges, but $a\s L$ remains finite (at least initially).

                     Now suppose we introduce a quartic nonlinear interaction, as just one example, leading
                     to the classical, phase-space, action functional
                       \b A=\tint\{\pi(t,\bx)\s{\dot\p(t,\bx)}-\half\s[\s\pi(t,\bx)^2+m_0^2\s\p(t,\bx)^2\s]\s
                       -g_0\s\p(t,\bx)^4\s\}\,dt\s d\bx \;. \e
                       Again, the temporal development of the field at one spatial point is independent
                       of the temporal development at any other spatial point.
                     Using the same lattice regularization as for the free theory and introducing
                     normal ordering for the interaction, the characteristic functional of the ground-state distribution is necessarily of the form
                       \b C(f)\hskip-1.3em&&=\lim_{a\ra0}N\int e^{\t (i/\hbar)\S_k\s f_k\s\p_k\,a^s-\S_k Y(\p_k,g_0,\hbar,a)\,a^s}\,\Pi'_k d\p_k\no\\
                       &&= {\cal N}\int e^{\t (i/\hbar)\tint f(\bx)\s\p(\bx)\,d\bx-\tint {\widetilde Y}(\p(\bx),g_0,\hbar)\,d\bx}\,
                       \Pi'_\bx d\p(\bx)\no\\
                       &&=e^{\t -(1/4\hbar\s{\sf m})\tint f(\bx)^2\, d\bx}\;.\e
                       Here $Y(\p_k,g_0,\hbar,a)$ and ${\widetilde Y}(\p(\bx),g_0,\hbar)$  denote some non-quadratic functions that arise in the solution of the Hamiltonian ground-state differential equation on the lattice and likewise in the formal
                       continuum limit, respectively. Importantly,  {\it the last line is a consequence of the Central Limit Theorem}, yielding a {\it free} theory with a mass {\sf m}, a factor that absorbs
                       all trace of the quartic interaction. In short, a conventional canonical quantization of this nonlinear (and nonrenormalizable) quantum field theory has rigorously led to a
                       ``trivial'' (free) theory, even though the original classical theory was nontrivial.

                       The question naturally arises: Can we change quantization procedures to yield a nontrivial result?

\subsection*{Affine quantization of ultralocal models}
                       We start by modifying the classical Hamiltonian before quantization. Namely,
                       for both $g_0=0$ and $g_0>0$, let us consider
                       \b H(\pi,\p)\hskip-1.3em&&=\tint\{\s\half[\pi(\bx)^2+m_0^2\s\p(\bx)^2]+g_0\s\p(\bx)^4\}\,d\bx\no\\
                       &&=\tint\{\s\half[\pi(\bx)\p(\bx)\p(\bx)^{-2}\p(\bx)\pi(\bx)+m_0^2\s\p(\bx)^2]
                       +g_0\s\p(\bx)^4\}\,d\bx\;.\e
                       On quantization, we will treat the classical product $\pi(\bx)\p(\bx)$ as the dilation function $\k(\bx)$ and invoke affine quantum commutation relations for which
                       $\k(\bx)\ra \hk(\bx)$ and $\p(\bx)\ra\hph(\bx)$ such that $[\hph(\bx),\hk({\bf y})]=i\hbar\s\delta(\bx -{\bf y})\s\hph(\bx)$. Note well: if we choose affine commutation relations for which $\hph(\bx)$ and $\hk(\bx)$, when smeared, are self-adjoint operators, then it follows that the canonical momentum operator $\hp(\bx)$, when smeared, is only a {\it form}  and {\it not} an operator due to the local operator product involved; recall that a form requires restrictions on kets {\it and} bras. In short, when quantizing fields, one must choose either affine variables or canonical variables, since both systems generally cannot exist simultaneously.

                          To see what affine quantization leads to, let us work formally and focus on twice
                          the kinetic energy density. Thus
                          \b \pi(\bx)^2\hskip-1.3em&&=\pi(\bx)\p(\bx)\p(\bx)^{-2}\p(\bx)\pi(\bx)\no\\
                      &&=\k(\bx)\p(\bx)^{-2}\,\k(\bx)\;, \e
                      which on quantization becomes
                      \b \hk(\bx)\s\hph(\bx)^{-2}\hk(\bx)\hskip-1.3em&&=\quarter[\hp(\bx)\hph(\bx)+\hph(\bx)\hp(\bx)]
                      \hph(\bx)^{-2}[\hp(\bx)\hph(\bx)+\hph(\bx)\hp(\bx)]\no\\
                      &&=\quarter[2\hp(\bx)\hph(\bx)+i\hbar\delta(0)]\hph(\bx)^{-2}[2\hph(\bx)\hp(\bx)-i\hbar\delta(0)]\no\\
                      &&=\hp(\bx)^2+i\s\half\hbar\s\delta(0)[\hph(\bx)^{-1}\hp(\bx)-\hp(\bx)\hph(\bx)^{-1}]\no\\
                      &&\hskip6em +
                         \quarter \hbar^2\delta(0)^2\hph(\bx)^{-2}\no\\
                         &&=\hp(\bx)^2+\threequarters\hbar^2\delta(0)^2\hph(\bx)^{-2}\;. \e
                         Note: the factor of $\threequarters$ will have an essential role to play.

                         Guided by this calculation we introduce the same lattice regularization to again quantize the classical free theory (i.e., $g_0=0$), but this time focussing on an affine quantization, namely
                         \b \H'_0=\half\S_k[-\hbar^2 a^{-2s}\d^2/\d\p_k^2+m_0^2\p_k^2+F\s\hbar^2 a^{-2s}  \p_k^{-2}-E_0\s]\,a^s\;, \e
                         where $F\equiv(\half-b\s a^s)(\threebytwo-b\s a^s)$; here $b>0$ is a constant factor with dimensions (Length)$^{-s}$, and $E_0$ is explained below. Note that $F$ is a regularized form of $\threequarters$, and becomes that number in the continuum limit. Again the ground state
                         of this Hamiltonian is a product over one-dimensional ground states for each
                         independent degree of freedom, and a typical ground state wave function, for small lattice spacing, has the form
                           \b \psi_k(\p_k)=(b\s a^s)^{1/2}\,e^{\t - m_0\phi_k^2\,a^s/2\hbar}\,|\p_k|^{-(1-2ba^s)/2}\;,\e
                           with a ground-state energy of $E_0=\half\s\hbar\s m_0 \s ba^s$; observe that the
                           ground-state energy on the lattice is finite  and that it{\it vanishes} in the continuum limit! This new form of ground-state wave function corresponds to what we have called a ``pseudofree''
                           model \cite{bookold}, and we see that it also arises upon quantization of the free classical model by exploiting an unconventional factor-ordering ambiguity and insisting on securing affine kinematical variables, $\hph(\bx)$ and $\hk(\bx)$.

                           The characteristic functional for the pseudofree ground-state distribution is given by
                           \b C_{pf}(f)\hskip-1.3em&&=\lim_{a\ra0}\int \Pi'_k\{ (ba^s)\,e^{\t i f_k\p_k\s a^s/\hbar-m_0\p_k^2\s a^s/\hbar}\,|\p_k|^{-(1-2ba^s)}\s\}\,\Pi'_k d\p_k\no\\
               &&=\lim_{a\ra0}\Pi'_k\s\{\s 1- (b\s a^s)\tint[1-\cos(f_k\p\s a^s/\hbar)]\,e^{\t -m_0\p^2\s a^s/\hbar}\,d\p/|\p|^{(1-2ba^s)}\s\}\no\\
                           &&= \exp\{-b\tint d\bx\tint[1-\cos(f(\bx)\s\l/\hbar)\s]\,e^{\t-b\s m\s\l^2/\hbar}\,d\l/|\l|\}\;,\e
                           where $m_0\equiv ba^s\s m$ and $\l\equiv\p\s\s a^s$; note the multiplicative renormalization involved in $m_0=b\s a^s\s m$. {\it No other number but $\threequarters$ would have led to  this desirable result!}

                           Finally, we consider the affine quantization of an ultralocal model with a quartic interaction as described by the classical action given above. In that case the lattice form of the quantum Hamiltonian acquires the additional term
                           $g_0\S_k\p_k^4\s a^s$ [with no normal ordering needed, but rather $g_0=(b\s a^s)^3\s g$] along with a suitable change of $E_0$. Now the lattice regularized
                           ground-state wave function at each site is of the form
                            \b \psi_k=(ba^s)^{1/2} \,e^{\t-\half\s u(\p_k,\hbar,a)\s a^s}\,|\p_k|^{-(1-2ba^s)/2}\e
                            for some suitable function $u(\p_k,\hbar,a)$. In turn, the characteristic functional of the interacting ground-state distribution becomes, in the continuum limit,
                            \b C(f)=\exp\{-b\tint d\bx\tint[1-\cos(f(\bx)\s\l/\hbar)\s]\,e^{\t -b^{-1}\s y(b\l,\hbar)}\,
                            d\l/|\l|\}\;, \e
                       where $u(\p,\hbar,a)$ is related to $b^{-1}y(b\l,\hbar)$ by $\l=\p\s\s a^s$ and multiplicative renormalization of suitable coefficients. General arguments from the theory of infinite divisibility show that only the free (Gaussian) or the nonfree (Poisson)
                       distributions, with characteristic functions for ground state distributions
                       illustrated above, are allowed by
                       ultralocal symmetry.

                       Note that following the route of affine quantization of the ultralocal model has resulted in overcoming the triviality of quantization that conventional canonical procedures invariably lead to. We also observe that suitable affine coherent states have been
                       central in the quantum/classical connection for ultralocal models; see \cite{bookold}.
                       Other approaches exist that lead to the same conclusions we have obtained here, but we have focussed on an affine variable approach since affine methods
                       are the principal theme of this article. This is the first time affine variables have
                       been introduced at the very beginning of the quantization procedure to lead, successfully, to the final solution for ultralocal models. As Voltaire might
                       have said: {\it Vive l'affine!}

                       It is also noteworthy that different but related methods have been used in a recent proposal to quantize {\it covariant} scalar fields in all spacetime dimensions. The reader interested in the application of noncanonical techniques to quantize covariant scalar models is advised to see \cite{klIOP}.

\section*{Affine Quantum Gravity}
Unifying classical gravity and quantum theory is a very difficult problem, and current approaches include
superstring theory and loop quantum gravity. Although these approaches have achieved much, there is still plenty of work to be done. In what follows we offer a sketch of the Affine Quantum Gravity program pioneered by the author. In a certain sense, quantum gravity is an ideal theory for an affine quantization approach!
\subsection*{Classical aspects of the affine variables}
In canonical phase-space coordinates, the classical theory of Einstein's theory of gravity in a  $3+1$ spacetime is described by the  action functional
   \b A=\int \{\s\pi^{a b}(x)\s{\dot g}_{ab}(x)-N^a(x)\s H_a(\pi,g)(x)-N(x)\s H(\pi,g)(x)\,]\,d^4\!x\;. \e
   Here $\pi^{ab}\s(=\pi^{b\s a})$ is the momentum, $g_{ab}\s(=g_{b\s a})$ is the metric, with $a,b=1,2,3$, and
   $N^a(x)$ and $N(x)$ are Lagrange multipliers to enforce, respectively, the diffeomorphism constraints  $H_a(\pi,g)(x)=0$ and
   the Hamiltonian  constraint $H(\pi,g)(x)=0$. The range of the momentum $\pi^{ab}(x)$ is all of ${\mathbb R}^6$,
   while the range of the spatial metric $g_{ab}(x)$ is such that the $3\times3$ matrix $\{g_{ab}(x)\}$
   is positive definite for all $x\in {\mathbb R}^4$; stated otherwise, if $u^a$ is an arbitrary, nonvanishing vector, then $u^a\s g_{ab}(x)\s u^b>0$. {\it Note well: The focus on positive definiteness of the spatial matrix $\{g_{ab}(x)\}$ for all $x$ is a fundamental principle of
   the affine formulation of both classical and quantum gravity}.
   This criterion implies the existence of the inverse metric tensor $g^{bc}(x)$, also positive definite, for which $g_{ab}(x)\s g^{bc}(x)=\de^c_a$. The classical equations of motion follow from stationary variation of the action and they can be expressed in terms of Poisson brackets as
    \b &&{\dot g}_{ab}(x)=\{g_{ab}(x), H_T\}\;, \hskip3em {\dot \pi}^{ab}(x)=\{\pi^{ab}(x), H_T\}\;,\no\\
       &&\hskip2em H_a(\pi,g)(x)=0\;, \hskip3em H(\pi,g)(x)=0\;. \e
       Here $H_T=\tint [\s N^a(x)\s H_a(\pi,g)(x)+N(x)\s H(\pi,g)(x)\s]\,d^3\!x$, and the
       fundamental Poisson bracket among the canonical variables at equal times is given by
       \b \{\s g_{ab}(x), \pi^{cd}(y)\}=
       \half[\s\de^c_a\s\de^d_b+\de^d_a\s\de^c_b\s]\,\de(x,y)\;. \e

       As dictated by the constraints, some aspects of the 12 phase-space variables at each point $x$ are unphysical.
       The Poisson brackets of the constraints among themselves all vanish on the constraint hypersuface, i.e,  that region of phase space defined by ${\cal C}\equiv\{(\pi,g): \s H_a(\pi,g)(x)=0,\s H(\pi,g)(x)=0\s\}$, and which are then classified as first-class constraints. It follows
       that once the
       phase-space variables $\pi^{ab}(x)$ and $g_{ab}(x)$ initially lie on the constraint hypersurface,
        then, as time passes, the phase-space variables remain on the constraint hypersurface for any
       choice of the Lagrange multipliers. Stated otherwise, the equations of motion do not determine the
       Lagrange multipliers, which is physically natural since the multipliers set the temporal coordinate distance to the next spatial surface as well as the location on that surface of the spatial coordinates themselves, all variables that may be freely specified. The constraints do not, however, form a (formal) Lie algebra since the structure ``constants'' are
       actually functions of the phase-space coordinates, a situation referred to as open first-class constraints. The constraints play a crucial role for gravity as they determine all the relevant physics;
       without them---imagine setting $N^a=N=0$ in the gravity action functional---there is not much physics left!

       Unlike the case of scalar field quantization, the dilation variable suitable for classical relativity is a familiar quantity, namely, $\pi^a_c(x)=\pi^{ab}(x)\s g_{bc}(x)$; it is also very easy to retrieve the original canonical momentum since $\pi^{ac}(x)=\pi^a_b(x)\s g^{b c}(x)$. The set of Poisson brackets satisfied by the classical affine variables $\pi^a_b(x)$ and  $g_{ab}(x)$
       at equal times is given by
          \b &&\{\pi^a_b(x),\pi^c_d(y)\}=\half[\de^c_b\,\pi^a_d(x)-\de^a_d\,\pi^c_b(x)]\,\de(x,y)\;,\no\\
  && \hskip-.08cm\{g_{ab}(x),\pi^c_d(y)\}=\half[\de^c_a\,g_{db}(x)+\de^c_b\,g_{ad}(x)]\,\de(x,y)\;,\\
  && \hskip-.18cm\{g_{ab}(x),g_{cd}(y)\}=0\;,\no  \label{e1}\e
  which follow directly from the fundamental Poisson bracket between $g_{ab}(x)$ and $\pi^{cd}(y)$
  given above. As was previously the case, the Poisson brackets for the affine variables form a (formal) Lie algebra.
  And just like the earlier studies of a single degree of freedom, one can reformulate the classical theory
  of gravity entirely in terms of these affine variables. Note well that $\pi^a_b$ has {\it nine}
  independent components while $g_{ab}$ has {\it six} independent components. Thus, there is no way in which the dilation variables $\pi^a_b$ can be considered the ``dual'' of the metric variables $g_{ab}$.
  To see the ``dilation'' aspects offered by $\pi^a_b$, consider the macroscopic canonical transformation induced by the arbitrary, nine-component function $\g^a_b(y)$ coupled with $\pi^b_a(y)$ in the manner
  $\pi(\g)\equiv\tint\s\g^a_b(y)\s\pi^b_a(y)\,d^3\!y$ as the generator, namely
      \b &&\hskip-2em e^{\t\s\{\,\cdot\, ,\s\pi(\g)\s\}}\;g_{cd}(x)\no\\
       &&\equiv g_{cd}(x)+\{g_{cd}(x),\pi(\g)\}+\half \{\{g_{cd}(x),\pi(\g)\},\pi(\g)\}+ \cdots\no\\
       &&\equiv M^a_c(x)\, g_{ab}(x)\s M^b_d(x)\;, \e
        where the matrix $M(x)$ with components
         \b M^a_b(x)\equiv\{e^{\t \half\g(x)\s}\}^a_b\e
           while $\g(x)$ is the $3\times3$ matrix $\{\g^a_b(x)\}$. Clearly, if $\{g_{ab}\}$ is a positive definite matrix
         before the transformation, it remains a positive definite matrix after the transformation. Even
         though the word ``dilation''does not capture all the properties of the matrix $M^a_b(x)$, we shall still refer to the affine variable $\pi^a_b(x)$ as the dilation variable. [{\bf Remark:} Occasionally,  we have used the word ``momentric''
         in referring to $\pi^a_b(x)$, a name that was derived from the two words {\it momen}tum and me{\it tric}.]

         A macroscopic canonical transformation with $\pi(\chi)\equiv \tint
         \chi_{ab}(y)\s\pi^{ab}(y)\,d^3\!y$ as generator would lead to $g_{cd}(x)+\chi_{cd}(x)$, a result that
         may transform a positive definite metric into a non-positive definite metric, thus potentially violating
         our fundamental principle.

         \subsection*{Quantum aspects of the affine variables}
         Our discussion of an affine quantization of gravity focusses on the use of affine variables for which $\pi^a_b(x)\ra\hp^a_b(x)$ and $g_{ab}(x)\ra \hg_{ab}(x)$, where these local operators are required to fulfill the affine commutation relations given by \b &&[\hp^a_b(x),\hp^c_d(y)]=i\half\s\hbar\s[\de^c_b\,\hp^a_d(x)-\de^a_d\,\hp^c_b(x)]\,\de(x,y)\;,\no\\
  &&\hskip-.1em[\hg_{ab}(x),\hp^c_d(y)]=i\half\s\hbar\s[\de^c_a\,\hg_{db}(x)+\de^c_b\,\hg_{ad}(x)]\,\de(x,y)\;,\\
  &&\hskip-.3em[\hg_{ab}(x),\hg_{cd}(y)]=0\;,\no  \label{e1}\e
  which are seen to be a direct transcription of the Poisson brackets for the classical affine variables. Observe that the affine commutation relations have the appearance of current commutation relations, and thus there are
  generally very different representations than arise in the case of variables that satisfy canonical commutation relations.

  The choice of a representation for the affine operators is largely determined by choosing a suitable fiducial vector for a set of coherent states. In our case, we introduce
    \b |\pi,\g\>\equiv e^{\t (i/\hbar)\tint \pi^{ab}(y)\s\hg_{ab}(y)\,d^3\!y}\,e^{\t -(i/\hbar)\tint \g^a_b(y)\s \hk^b_a(y)\,d^3\!y}\,|\eta\>\;, \e
    and look for arguments to choose a suitable fiducial vector.
    To proceed further, we need to introduce some physics. Dirac's quantization procedure for systems with constraints requires quantization first and reduction second. In brief, we are asked to quantize initially as if there were no constraints at all; after quantization, we are then required to restrict the original Hilbert space $\frak H$ to the physical Hilbert space ${\frak H}_{phys}$ by some means.
    Therefore, we initially focus on a quantization without regard for the constraints, which means taking a neutral position regarding how field values at one spatial point are related to field values at any other spatial point. The neutral and natural way to do so is to assume that the initial operator representation is ultralocal in character. This means, for example, that the proper  coherent state overlap function
    must initially have the form
      \b \<\pi'',\g''|\pi',\g'\>\equiv e^{\t-\tint L[\pi''(x),\g'(x);\pi'(x),\g'(x)]\,d^3\!x} \;.\e
      If we imagine regularizing this expression by a spatial lattice, then we can find the form of $L$ by studying what happens at a single lattice site. This study has been carried out in \cite{aqg1} by
      making a choice of the fiducial vector that---like the elementary one-dimensional affine example studied earlier---is annihilated by a complex linear combination of the affine variables (i.e, the fiducial vector becomes an extremal weight vector). The result of that study leads to
      \b  \<\pi'',\g''|\pi',\g'\>\hskip-1.3em&&= \exp\bigg(\!-\!2\int b(x)\,d^3\!x\, \no\\
  &&\hskip-3em\times\ln\bigg\{  \frac{
\det\{\half[g''^{kl}(x) +g'^{kl}(x)]+i\half [\hbar\s b(x)]^{-1}[\pi''^{kl}(x)-
\pi'^{kl}(x)]\}} {(\det[g''^{kl}(x)])^{1/2}\,(\det[g'^{kl}(x)])^{1/2}}
\bigg\}\bigg)\no\\
  && \equiv\<\pi'',g''|\pi',g'\> \;.\label{e3}\e
Observe that the matrices $\gamma''$ and $\gamma'$ do {\it not} explicitly appear in this expression; the
 choice of $|\eta\>$ is such that the matrix $\{\gamma^a_b\}$ has  been replaced by the
 positive definite matrix $\{g_{ab}\}$, where
  \b  g_{ab}(x)\equiv M_a^c(x)\,\<\eta|\hg_{cd}(x)|\eta\>\,M_b^d(x)\;, \e
      and, as before, the matrix elements $M_a^c(x)=\{e^{\t\half\gamma(x)}\}_a^c$. Symmetry of the fiducial vector has reduced  dependence of the coherent states from the nine components of $\g^a_b$ to the six components of $g_{ab}$, and therefore we are free to relabel the coherent states as $|\pi,g\>\equiv |\pi,\g\>$ to reflect that symmetry. That is why we have already labeled the coherent state overlap $\<\pi',g''|\pi',g'\>$ as well. In this expression,
      the function $b(x)>0$ is a scalar density with dimensions (Length)$^{-3}$, and thus we observe that the coherent state overlap function is {\it invariant} under arbitrary spatial coordinate transformations.
      Moreover, the physical meaning of the affine coherent state labels $\pi^{ab}(x)$ and $g_{ab}(x)$ follows from the fact that
         \b \<\pi,g|\s\hg_{ab}(x)\s|\pi,g\>\hskip-1.3em&&=g_{ab}(x)\;, \no\\
            \<\pi,g|\s\hp^a_c(x)\s|\pi,g\>\hskip-1.3em&&=\pi^{ab}(x)\s g_{bc}(x)\equiv \pi^a_c(x)\;. \e
      The coherent state overlap is a function of positive type, and therefore it can serve as a reproducing kernel for a natural functional representation of the original Hilbert space $\frak H$ by continuous functionals \cite{aron}.

      Much as the affine coherent state overlap function for a single degree of freedom admitted a path integral formulation with a Wiener measure regularization (by considering the propagator when the Hamiltonian vanishes), it is also true
      that the affine coherent state overlap function for gravitational affine variables is given by
      \b  && \<\pi'',g''|\pi',g'\>  \no\\
 &&\hskip1cm=\lim_{\nu\ra\infty}{\o{\cal N}}_\nu\s\int
e^{\t-(i/\hbar)\tint[g_{ab}\s\s{\dot\pi}^{ab}]\,d^3\!x\,dt}\no\\
  &&\hskip-1cm\times\exp\{-(1/2\nu)\tint[\s\s[\hbar\s b(x)]^{-1}g_{ab}\s g_{cd}\s
{\dot\pi}^{bc}\s{\dot\pi}^{da}+[\hbar\s b(x)]\s g^{ab}\s g^{cd}\s{\dot g}_{bc}\s{\dot g}_{da}\s]\,
d^3\!x\,dt\}\no\\
  &&\hskip2cm\times[\Pi_{x,t}\,\Pi_{a\le b}\,d\pi^{ab}(x,t)\,
dg_{ab}(x,t)]\;. \label{e8} \e
   Note well: Although this functional integral has all the appearances of a phase-space functional integral, it is important to  keep in mind that this formulation arose from an {\it affine} quantization procedure and used
    {\it affine} coherent states.

      The reader is encouraged to compare this functional integral with the Wiener measure regularized path integral for affine coherent states applied to a single degree of freedom with a vanishing Hamiltonian, and especially to note the formal similarity of the two metrics that serve to define the Brownian motion and which also ensure that the required positivity conditions on suitable variables are maintained throughout the functional integral.

    So far we have not included the effect of the constraints in accordance with
    the Dirac approach to the quantization of systems with constraints \cite{dirac}, where the rule is to quantize first and reduce second. The reason behind this rule is that by reducing first, there is the strong chance that the remaining classical phase space does not admit Cartesian coordinates leading to an
    uncertain quantization procedure. By choosing to quantize first, one can usually
    arrange a physically secure quantization procedure. After the initial step of quantization, Dirac advocates generating the physical Hilbert space from the original Hilbert space by asking that the constraint operators all give zero when acting on those vectors that make up the subset of
    the original Hilbert space that forms the physical Hilbert space. This procedure works for constraints
    that form elements of a compact Lie algebra. However, this procedure encounters problems with constraints for which zero lies in the continuous spectrum as well as for all constraint systems that are second class (defined by Poisson brackets or commutators of constraints that do not vanish on the constraint hypersurface or on the physical Hilbert space, respectively). A more recent variation of Dirac's procedure  \cite{proj} instead builds a projection operator $\E$ that projects onto a small spectral subspace, between zero and $\delta(\hbar)^2$ (where  $\de(\hbar)$ is a regularization parameter), of the sum of the squares of the constraint operators, a method that treats both first- and second-class constraints on an equal footing. The regularized  physical Hilbert space (regularized by $\de(\hbar)$) is then
     defined by ${\frak H}_{phys}=\E\s\s{\frak H}$. To enforce the constraints properly, it is necessary to reduce the regularization parameter $\de(\hbar)$ to its smallest optimal size, which can either be zero (for first-class constraints) or nonzero (for second-class constraints). If the spectrum of the sum of the squares of the constraint operators near zero lies in the continuum, then it is necessary to take a suitably rescaled  form limit of the projection operator $\E$ as $\de\ra0$. Advantages of this method include:
     no gauge fixing, a path integral formulation with no auxiliary variables, no need for Dirac brackets or other means to eliminate second-class constraints, as well as several others. Since quantization of constraints is not a primary topic of this article, we have only offered a very brief summary of the projection operator method of dealing with systems with constraints; the interested reader can find the full story of this method in \cite{proj,proj2,sch,agq2}.

     Even though we have offered few details, we have sketched enough information so that the reader
     will be able to sense the meaning of an introduction of the constraints in a regularized fashion.
     In this discussion, we seek the affine coherent state matrix elements of the regularized projection operator $\E$ by means of a specially weighted functional integral involving the Lagrange multiplier functions,
     $N^a(x)$ and $N(x)$, and the constraint symbols, ${\tilde H}_a$ and ${\tilde H}$, (symbols because $\hbar>0$), along with a suitable measure, $R(N^a,N)$. The resultant functional integral is formally a modest generalization of the functional integral for the affine coherent state overlap, and is given by
      \b  && \<\pi'',g''|\s\E\s|\pi',g'\>  \no\\
 &&\hskip1cm=\lim_{\nu\ra\infty}{\o{\cal N}}_\nu\s\int
e^{\t-(i/\hbar)\tint[g_{ab}\s\s{\dot\pi}^{ab}+N^a\s{\tilde H}_a+N\s{\tilde H}]\,d^3\!x\,dt}\no\\
  &&\hskip-1cm\times\exp\{-(1/2\nu)\tint[\s[\hbar\s b(x)]^{-1}g_{ab}\s g_{cd}\s
{\dot\pi}^{bc}\s{\dot\pi}^{da}+[\hbar\s b(x)]\s g^{ab}\s g^{cd}\s{\dot g}_{bc}\s{\dot g}_{da}]\,
d^3\!x\,dt\}\no\\
  &&\hskip2cm\times[\Pi_{x,t}\,\Pi_{a\le b}\,d\pi^{ab}(x,t)\,
dg_{ab}(x,t)]\,\D R(N^a,N)\;. \label{e8} \e
This expression can serve as a reproducing kernel for a functional representation of the regularized physical Hilbert space ${\frak H}_{phys}$ by continuous functionals.
For further details regarding the program of affine quantum gravity, the interested reader is referred to
\cite{slavnov} and references therein.

\section*{Summary}
In this article we have tried to persuade the reader that there are many useful similarities between canonical variables and affine variables, both classically and quantum mechanically. In particular, we have stressed
that for finitely many degrees of freedom, one may pass from canonical to affine variables and vice versa. Moreover, for the purpose of quantization,  the value of Cartesian canonical variables is essentially matched by the utility of affine variables defined in terms of those canonical variables. After the analysis of a toy problem using both a direct canonical quantization and a direct affine quantization, we
examined a scalar field theory and showed how a direct canonical quantization fails to provide satisfactory results for a nonrenormalizable scalar quantum field theory, while, importantly, a direct affine quantization of the same model leads to completely satisfactory results. Finally, we explained how affine variables and affine coherent states can play a profound role in studying quantum gravity, a program that, due to its
extreme complexity, is still far from complete.

\end{document}